\documentclass{article}
\usepackage[utf8]{inputenc}

\title{Patent-publication pairs for the detection of knowledge transfer from research to industry: reducing ambiguities with word embeddings and references}
\author{
       Klaus Lippert$^1$ and Konrad U. Förstner$^{1,2}$\\ 
       $^1$\begin{small}ZB MED - Information Centre for Life Sciences  \end{small}\\
       $^2$\begin{small}TH Köln – University of Applied Sciences, Institute of Information Science \end{small}
       }

\date{December 1, 2024}
\usepackage{fancyhdr}  
\usepackage{natbib}
\usepackage{graphicx}
\usepackage{color}
\usepackage[utf8]{inputenc}
\usepackage{caption}
\usepackage[a4paper,  margin=1.5in]{geometry}
\usepackage{natbib,hyperref}

\begin{document}

\maketitle

\section*{Abstract}
The performance of medical research can be viewed and evaluated not only from the perspective of publication output, but also from the perspective of economic exploitability. Patents can represent the exploitation of research results and thus the transfer of knowledge from research to industry. In this study, we set out to identify publication-patent pairs in order to use patents as a proxy for the economic impact of research. To identify these pairs, we matched scholarly publications and patents by comparing the names of authors and investors. To resolve the ambiguities that arise in this name-matching process, we expanded our approach with two additional filter features, one used to assess the similarity of text content, the other to identify common references in the two document types. To evaluate text similarity, we extracted and transformed technical terms from a medical ontology (MeSH) into numerical vectors using word embeddings. We then calculated the results of the two supporting features over an example five-year period. Furthermore, we developed a statistical procedure which can be used to determine valid patent classes for the domain of medicine. Our complete data processing pipeline is freely available, from the raw data of the two document types right through to the validated publication-patent pairs.
 
\section*{Introduction}
Nowadays, research performance in the medical sciences is mainly evaluated on the basis of publication output (i.e. articles in scholarly journals) and the acquisition of external funding. This is applied at all aggregation levels: from individual scientists to faculties and institutions and all the way up to entire countries. 
However, scientific impact is also reflected in many other parameters, such as, for example, the training of young scientists, lecturing activities, participation in committees, and the development of new processes and products.
Evaluating research performance more holistically requires additional performance measures. This has prompted a great deal of interest in efforts to identify and measure the transition of knowledge into specific areas, such as health policy recommendations (\cite{zbw_altmetrics}) and other policymaking domains, and the broader translation of knowledge into the economy.
The presented work traces the diffusion of knowledge into the economy by linking patents to underlying scholarly articles and their authors.

Identifying patents that stem from research and matching them to the corresponding academic publications is question of interest, especially when it comes to evaluating the innovative capacity of institutions. A simple starting point is the use of employee lists (e.g. \cite{VANDONGEN201427}) or specially created patent databases (e.g. the KEINS database, see \cite{lissoni1}) that allow users to specifically search for patents from an academic environment. The KEINS database does not provide personalised data such as names or patent numbers, however, so it cannot be applied to the evaluation of the workflow presented here. 
The process of matching publications to patents involves a lot of manual work, both in collecting the initial data and in resolving ambiguities caused by homonyms. The resolution of ambiguities is accomplished through online surveys, for example, or even through personal contact (e.g. \cite{sweden1}). This requires significant effort, which is why investigations are usually limited to one (e.g. \cite{mit1}) or a few (e.g. \cite{VANDONGEN201427}) institutions. The next generalising step away from manual work and towards automation is the use of publications and thus the author's name as a starting point. For example, in their large-scale approach, \cite{Dornbusch2015Academic} filter identified publication-patent pairs in order to reduce ambiguities. The present work follows this approach (and the use of those filters) and extends it with two new filter approaches. Both these approaches ‒ comparing content through embedding, and using statistical methods to restrict the permitted patent classes ‒ are innovations in the field of publication-patent matching. Furthermore, by publishing the code of our workflow, we enable others to reproduce the results of the data example described in this paper and build on it in and further work, such as e.g. embeddings with the latest vector spaces.

Something all the previous works mentioned above have in common is the way in which they aggregate, identify and validated publication-patent pairs at the institute or country level in order to compare variables such as innovation capacity. The focus of the present work is to improve the matching of individuals and their publications or patents using the new filters.

The document is organized as follows: first, we describe the publication and patent data used in this study. Next, we outline the workflow developed for this data, in which the newly introduced filter features play a crucial role. After describing the matching of publications and patents based on common names, we present and examine these new features in detail and discuss their use as filters for the raw patent-publication pairs identified in the previous steps. We then show how these results are confirmed by a small-scale manual evaluation. This paper concludes with a brief discussion of the results and the future outlook.

\section*{Data}
\label{datengrundlage}
To keep the presented study FAIR\footnote{Findable, Accessible, Interoperable, and Reusable (see e.g. \cite{fair1})}, two open data sets are used: patent data from the European Patent Office (EPO)\footnote{www.epo.org $\rightarrow$  data.epo.org/publication-server/rest/v1.2/publication-dates} and publication data from MEDLINE/PubMed (\cite{pmc}) for publications. Additionally, all the code (SQL and Python) required to reproduce the presented analysis and figures is openly available\footnote{https://github.com/foerstner-lab/patent$_-$publication$_-$links  DOI: https:/doi.org/10.5281/zenodo.13354449}.

\begin{figure}[h!]\centering
\captionsetup{width=1\linewidth}
\includegraphics[width=1\linewidth]{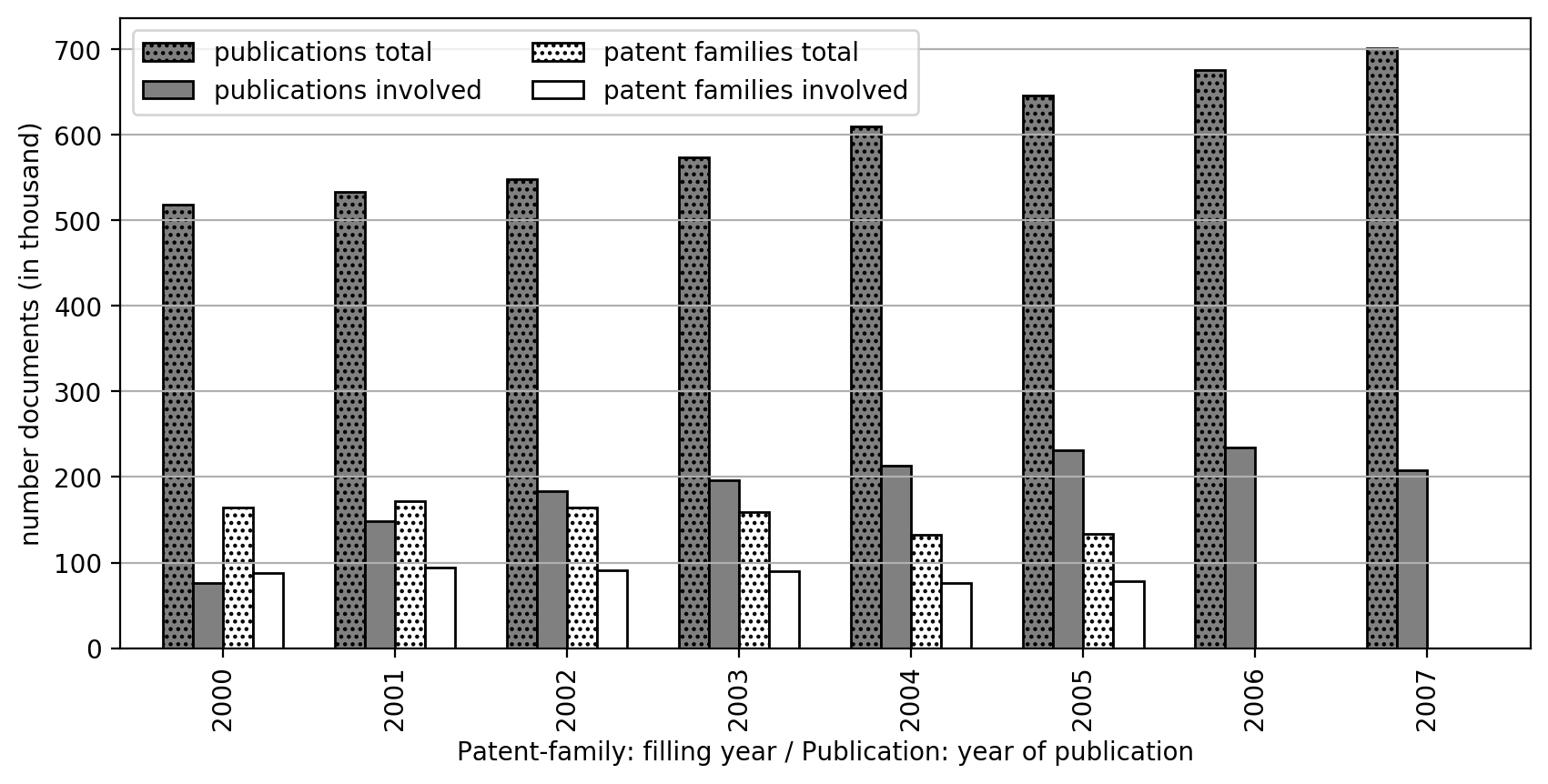}
\caption{Number of patent families from EPO and publications from PubMed baseline dataset for the period of this study, plus number of each document type involved in the raw publication-patent pairs. }\label{fig_available_datasets} 
\end{figure}

To keep the amount of data manageable for reproduction, this study is limited to a five-year period of patent first filing dates. As described in the later sections, patent data from later years are also taken into account by grouping the patents into so-called patent families. The other data source, publications, is drawn from the same period plus an additional two years subsequent to the five-year period. This takes into account the fact that only previously unpublished content can be patented (i.e. content that does not constitute "prior art"). This is also described in detail later in this paper. The years 2000 to 2005 were chosen as the period in which the first patent filing was made.
This period was chosen because the number of patents in this period was still manageable and many of the publications were already available in digital form in the PubMed data set. Figure \ref{fig_available_datasets} gives an overview of the orders of magnitude of the raw data used.

Patent data contains not only the inventors but also the applicants of the patent, who later ‒ once the patent is granted ‒ become the legal owners of a patent. Inventors are usually natural persons, whereas applicants can be individuals, companies or scientific institutions. In the first decade of this century, some European states changed their laws: for example, in Germany, a professor was previously allowed to be registered as an applicant for a patent application (\cite{wiki:hochschullehrerprivileg}), but after 2002 this professors' privilege was abolished and the scientific institution had the right to be registered as an applicant. This has made scientific patents more easily recognisable in the case of Germany, for example. Other European countries already had this regulation in 2002 or implemented similar ones over the years. 
We did not investigate the impact of professor's privilege in any greater detail; this has already been examined, for example, by \cite{provprivi2} for Denmark and by \cite{profprivi} for Germany.

\section*{Data processing}   
\begin{figure}[h!]\centering
\captionsetup{width=1\linewidth}
\includegraphics[width=1\linewidth]{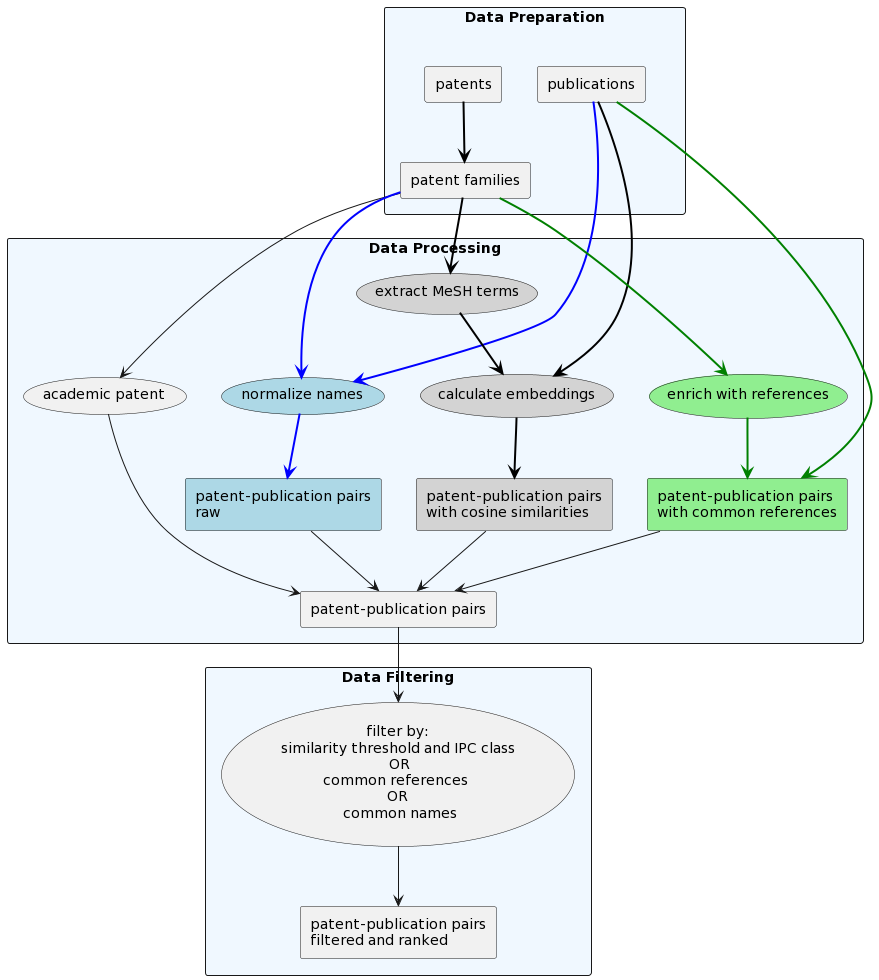}
\caption{Simplified workflow}\label{fig_workflow_simple} 
\end{figure}

Figure \ref{fig_workflow_simple} shows the simplified workflow used in this study to assign patents to corresponding publications. The workflow can be roughly divided into three successive work blocks (blue boxes in Figure \ref{fig_workflow_simple}): (1) data preparation, (2) data processing and (3) data filtering. 
The main part of the workflow is data processing. This step consists of (a) finding raw patent-publication pairs with the same author/inventor names (blue path in Figure \ref{fig_workflow_simple} ) and then building on this path by creating three parallel paths that use additional supporting features to verify the identified raw patent-publication pairs: 
(b) content similarity using cosine similarity of documents in a word vector space (grey path), (c) common references (green path) and (d) indicators for an academic patent (light grey path).
In the following sections, each step of this workflow is described and each finding is examined in detail.

\subsection*{Patent families}
Based on the process a patent goes through from the first filing with the patent office to the eventual granting of the patent and assignment of the identification number by the patent office, patent data can be split into two main groups: patent applications (second last digit of patent's publication number is A) and patent specifications (second last digit is B). Within each group, the final digit indicates the chronological order of filing. 
\\
A patent family is the totality of all patents filed for an invention, regardless of which patent office and which country they were filed in. Thus, the same invention may have been assigned different patent numbers by different patent offices.
In this study, only patents published by the European Patent Office are grouped into patent families. All these patents start with the letters "EP". In the following sections, patent families are defined as patent publications that were assigned the same patent number (except for the last letter A or B and the following chronological number) by the European Patent Office. \footnote{e.g. EP00100008A1, EP00100008A2 and EP00100008B1 would be grouped together in one patent family with EP00100008}
This grouping affects nearly all data fields in a patent data set, in particular inventors, applicants, titles, description and claim texts, etc. The filing date of the first patent within a family is taken as the filing date of the whole patent family.
\\
Grouping patents into families also allows us to consider data sets from outside the 
chosen study period. For example, a patent family with a first filing year of 2001 might include documents published in 2010, which is outside the chosen study period (see below).

\subsection*{Pairing publications and patents using common names}
Patents and publications are paired by matching inventor and author names. There are slight differences in the quality of the data in these two data sets, and consequently in how they are processed. 
Both sets consist of last names and a mixture of full first names and just the initials. Umlauts and other special characters are replaced during processing. In addition, the patent inventor data field often includes academic titles such as PhD, Dr, MBA, Professor, etc., which are also removed. In the publication author data fields, the first, middle and last names are stored separately, unlike the inventor data field, which has no such restrictions and can thus accommodate any conceivable way of writing a name. The task of normalising names is conducted by a Python library \footnote{https://pypi.org/project/personnamenorm \ \ \ DOI:10.5281/zenodo.11182093} created in this project, which tackles name ordering on the basis of rules and an additional Naive Bayes approach.
The resulting normalised names have the form "last name, initials of first names". 
We are aware that this normalisation increases the number of 
homonyms (i.e. different persons with identical names) but the use of this least common denominator for the names in our data sets is a common approach. According to the simulations of \cite{Milojevic2013AccuracyOS}, using only the first initials already correctly identifies 97$\%$ of authors. More advanced methods for author name disambiguation using more metadata than just the name can be found e.g. in \cite{ferreira1} or \cite{jinseok1}.

After normalisation, possible duplicate \footnote{The very unlikely case of the same first and last name may indeed occur, though normalisation makes identical names more likely. For example, the author of this study, Klaus Lippert, and Karl Lippert would both be normalised to "K Lippert". } names in one publication or one patent family respectively are removed. This deduplication prevents the overestimation of $n^2$ connections when applying an SQL JOIN statement in the process of building patent-publication pairs.  

One of the main legal requirements for a patent is novelty. In our context of publication-patent matching, this means the filing date of the patent must be before the publication date of the publication. To reduce computational costs, 
for each filing year of the patents only publications from the same or the two consecutive years are included in the SQL JOIN statement \footnote{e.g. patent filling year 2000 is joined with publications from year 2000, 2001 and 2002}. After that, the obtained pairs are filtered by date to satisfy the criterion of novelty. We follow here the argumentation of \cite{Dornbusch2015Academic}, who set a minimum time interval of half a year for the duration of the publication process. We limit the time interval between 0.5 and 1.5 years.

In regard to geolocation, the patent data contain the addresses of the inventors, the publication data the affiliations of the authors. To tackle the multiplicity of identical names, a further restriction must be placed on pair formation: \cite{Dornbusch2015Academic} only allow geographical distances of max. 30 km between affiliation and residence of the author. An improvement at this point would be to replace geographical distance with the use of a "connectivity" (see \cite{Brockmann2017Global}), which includes e.g. fast travel routes from cities outside the max. geographical distance. Both these approaches are beyond the scope of this work, and we therefore restrict ourselves to the country level. For the publications, the country information is extracted either from existing email addresses or from the text of the affiliations and is then normalised to the two-digit ISO country codes used in the patents. If this geographical information exists only for the first author, it is transferred to all other authors. 
For the pairing (using a SQL JOIN statement), data sets with country information are treated preferentially.
Figure \ref{fig_available_datasets} gives an overview of how many patents and publications are involved in this raw matching.

The described procedure for pairing patents and publications leads to enormous numbers of many-to-many relationships, especially when only one or two names are involved. Therefore, the identified pairs are subsequently filtered, as described in the following sections. The progress of this filtering can be seen in figure \ref{fig_NN_relationsships}.

\subsection*{Patent-publication similarity scores using embeddings}

The main tool used in this study to reduce the enormous number of many-to-many relationships is a comparison of the content of patents and publications. The relevant content is condensed by using only words that are available in the well-known MeSH thesaurus (\cite{mesh_en}).

The condensed content is mapped to a word vector space and a single vector is derived for each document. 
With this numerical representation, the two documents of each identified patent-publication pair can be examined and compared to assess their similarity in content. For this purpose, the cosine similarity (see e.g. \cite{cosdist}) is used. In short, the cosine similarity ranges between 0 and 1, and in the context of comparing two vectors of word embeddings, 1 stand for identical documents and 0 for "no similarity at all". 

In the present study, the predictive-based \footnote{Context-predicting as opposed to context-counting, see \cite{dontcountpredict}} word vector space
 BERT (\cite{bert}) is used. In the following section, we describe the exact procedure for moving from raw text to similarity scores for the patent data.

\begin{figure}[h!]\centering
\captionsetup{width=1\linewidth}
\includegraphics[width=1\linewidth]{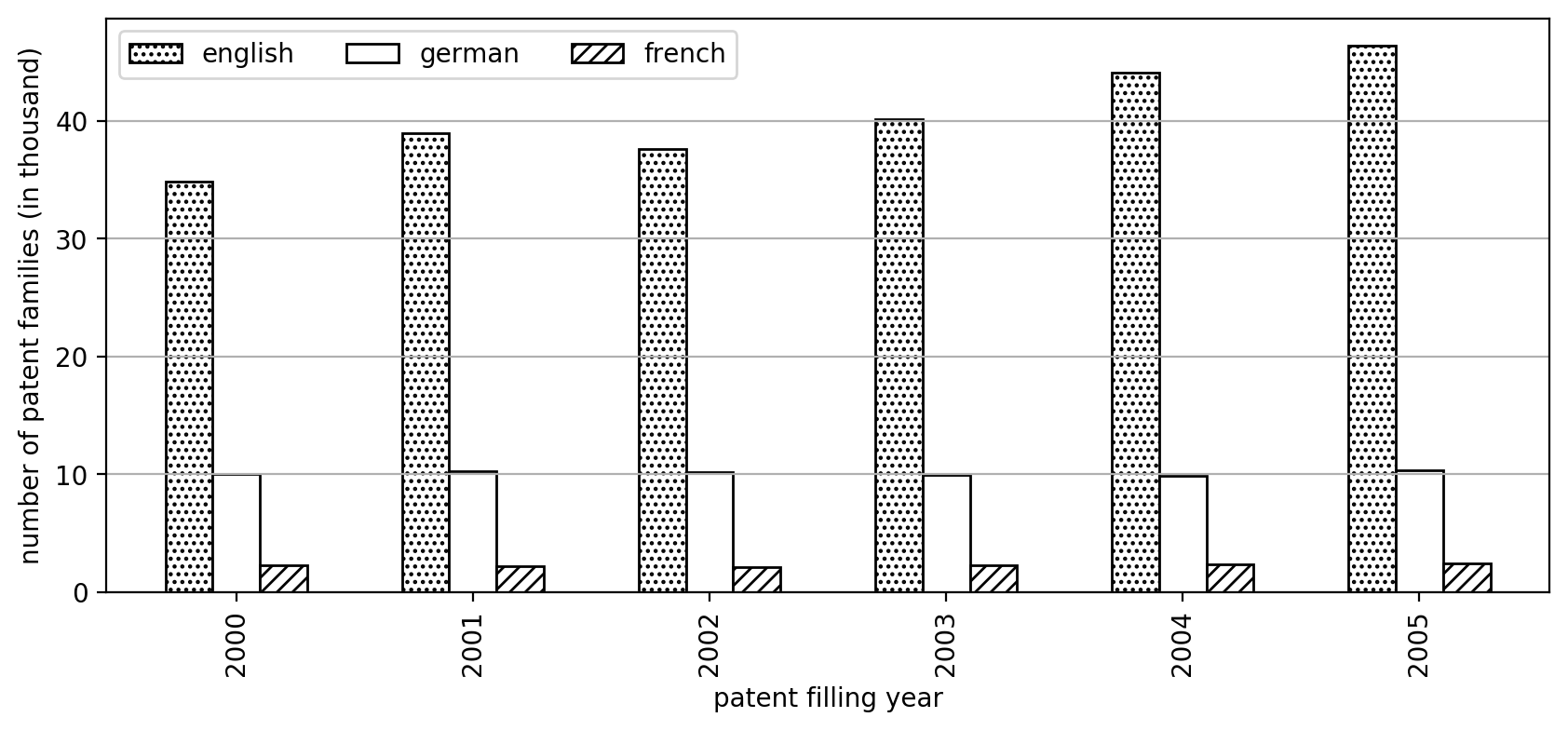}
\caption{Language of patent descriptions used for MeSH extraction.} \label{fig_languages_patent_description} 
\end{figure}

As a first step, the patent descriptions of the individual patent families are summarised and duplicate texts are removed. Patents at the EPO may be in three languages: English, German and French. The patent descriptions are ranked in this order if more than one language is available. The order English - German - French is chosen arbitrarily and corresponds to the experience of this study's authors in machine processing in the respective language. MeSH main headings and entry terms (= synonyms) are extracted from these texts.  This is carried out in the three languages using the English MeSH (\cite{mesh_en}) and the two official translations in German (\cite{mesh_de}) and French (\cite{mesh_fr}). 
The extraction is performed using a simple look-up procedure, which considers pure upper-case words (like for example the WHO) and  also main headings consisting of several words: First, the high n-grams are extracted and masked. This is repeated stepwise down to 1-grams.  To keep the computational expenses small, only the first occurrence of a word is considered.
Subsequently, all extracted German and French main headings and synonyms are mapped to their English main heading counterparts. Figure \ref{fig_languages_patent_description}  illustrates the distribution of the languages used: although English clearly predominates, German and French have a non-negligible share.
In the resulting excerpt of the patent description, duplicate words are now removed and sorted alphabetically. 
For each word of this new "text", a vector is created using the BERT base model and the sum of all vectors results in one vector per patent.

In the BERT model, there are (slightly) different vectors for every single word, depending on the context surrounding that word. The authors are aware that ordering the words confuses the original order. However, the clear results found (see below) justify this approach.

We are also aware that there are pre-trained BERT models that are better adapted to the life sciences domain (e.g. BioBert (\cite{Lee_2019}). Furthermore, at the time this study was conducted, Large Language Models were not widely used or available. However, the availability and rapid development in this area are promising in terms of improving the results of the presented approach. 
The focus of this work was to create an automated workflow and to explore the possibilities of comparing the content of these two text types, publications and patents. This will serve very well as a baseline for further work with state-of-the-art embedding models, both in terms of the results and reuse of the workflows.
For the sake of simplicity, however, we used the BERT base model, which provided is with satisfactory results. 

Another conceivable way of transforming the descriptions into a word vector space would be to use a graph2vec method \footnote{e.g. node2vec \cite{node2vec}, rdf2vec \cite{rdf2vec}, owl2vec \cite{owl2vec}}. In this case, the information on the graph structure of the MeSH ontology would flow into the word vector space. 

When "academic patents" are referred to in this paper, it is clear either from the inventor's name that the patent is held by a scientific employee or from the application name that the patent is held by a scientific institution. Both possibilities are described in detail at the appropriate point below. These apparent "academic patents" are, of course, only a subset of knowledge translation from academia to industry, but they provide valuable additional evidence for the research described here. 
\\
On the publication side, however, things are much simpler: the PubMed data contain human-curated MeSH main headings, which contain only the main topic of the text. The conversion to a vector per document is straightforward and analogous to the patent data.

Figure \ref{fig_languages_patent_description} shows the results of the similarity calculation of  identified patent-publication pairs plotted against the number of overlapping names. The boxes show the 25$\%$ and 75$\%$ range of the distribution, the whiskers mark the 5$\%$ and 95$\%$ range. It is easy to see that the cosine similarity for patent publication pairs with three or more common names is in the range greater than approx. 0.7. We consider this number of common names to be very unlikely to be random and thus regard it as a valid patent-publication pair. Using the median, a threshold is calculated from the lower bounds of the whiskers of the boxplots of two or more common names. This threshold is used later in this work to filter pairs and thus to reduce the many-to-many relationships.

\begin{figure}[h!]\centering
\captionsetup{width=1\linewidth}
\includegraphics[width=1\linewidth]{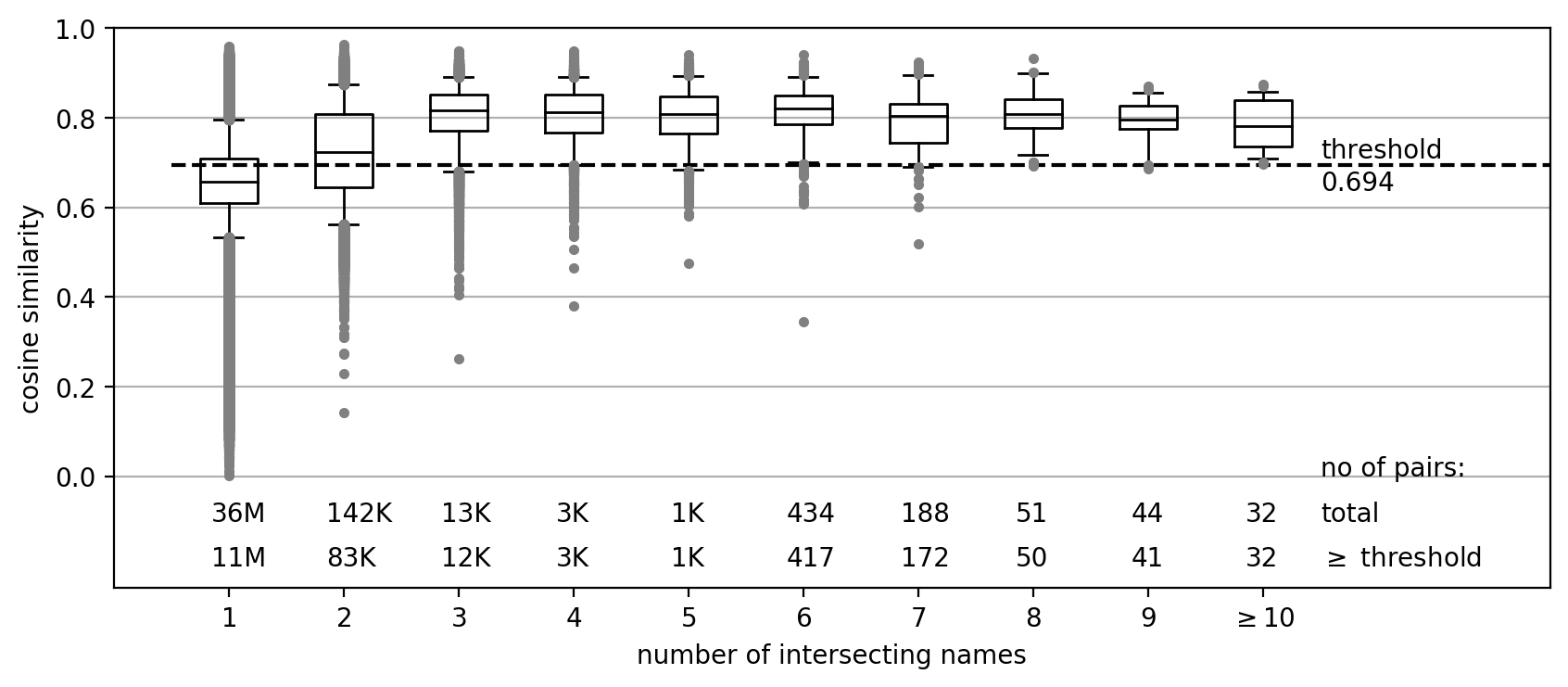}
\caption{Cosine similarities of patent-publication pairs separated by the number of common author/inventor names. Cosine similarity is derived from the BERT base model using MeSH main headings from patents and publications. In addition, the numbers of the respective patent-publication pairs are given for the respective numbers of common names. }
\label{fig_common_names_vs_cossim} 
\end{figure}

\subsection*{Common references for patent-publication pairs}

Another important tool for improving publication-patent matching and consequently reducing the number of many-to-many relationships is an analysis of common references in patents and publications, which indicate that both text types deal with the same topic. 
The idea of using common references to identify similar content has already been successfully applied to the problem of "prior art" by e.g. \cite{berndpaper}, who compared patents with patents.
\\
References appear in two places in a patent document: in the references section, and as in-text references. In the approach outlined here, we only consider references that appear in the 
references section. 
The additional in-text references are not strictly required for the focus set of this work. Harvesting these references is the subject of recent studies, e.g. \cite{Verberne2019ExtractingAM} and \cite{DBLP_journals/corr/abs-2101-01039}.
\\
For the references in the references section, the DOI was chosen as the identifier. On the publication side, this is already available in the PubMed dataset.
In the case of a patent, it is not mandatory to provide references and there is no standard citation style. Very few patents contain the DOIs of referenced publications. The patent references were therefore enriched with DOIs using the Crossref API. The API offers a free of costs search option which was used for title, authors, journal and publication year. To minimize false results, the three "best" results returned by the API were examined for exact matches and re-ranked accordingly, and the best one was chosen.

\begin{figure}[h!]\centering
\captionsetup{width=1\linewidth}
\includegraphics[width=1\linewidth]{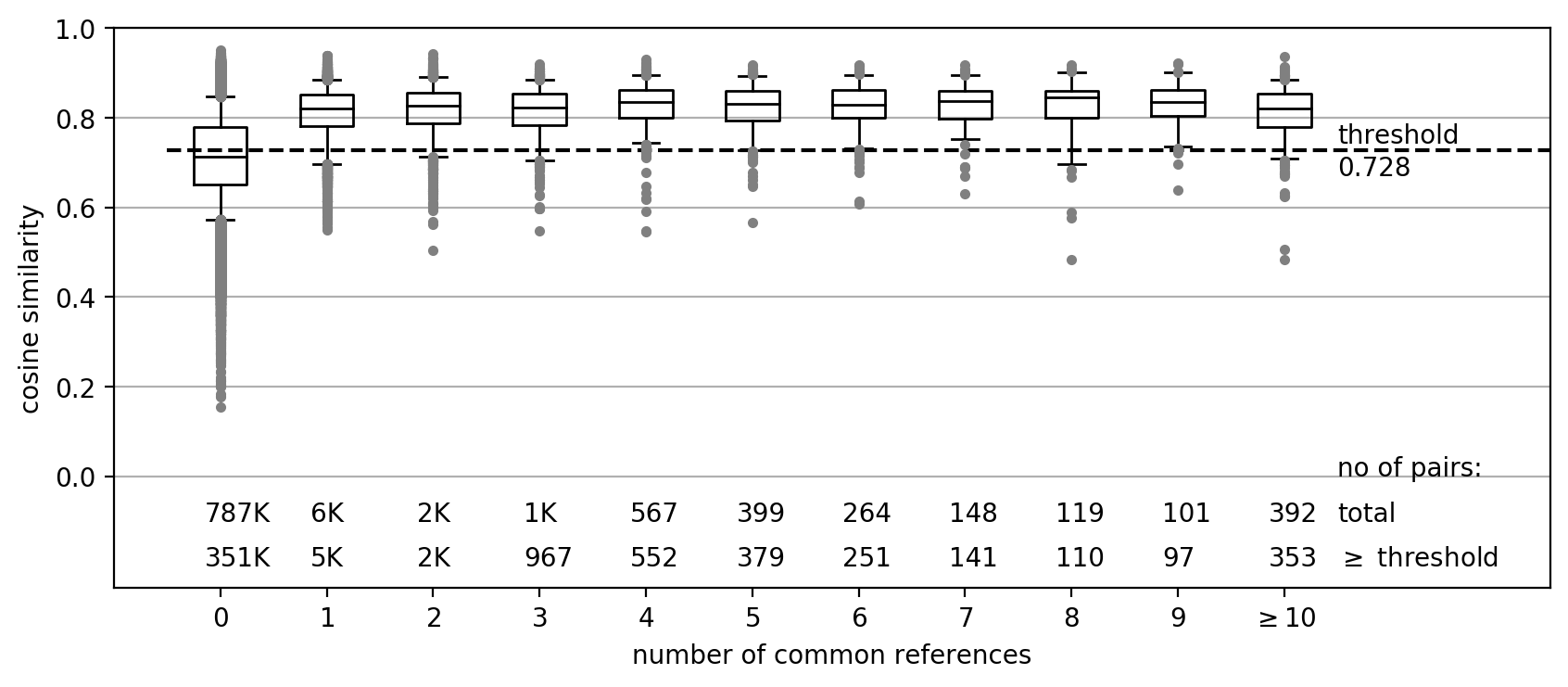}
\caption{Cosine similarities of patent-publication pairs separated by the number of common references. 
Cosine similarity is derived from the BERT base model based on MeSH main headings from patents and publications.
In addition, the numbers of the respective patent-publication pairs are given for the respective numbers of common references.}
\label{fig_common_references_vs_cossim} 
\end{figure}

Figure \ref{fig_common_references_vs_cossim} shows identified common references plotted against cosine similarity. 
The properties of the boxplot are the same as in Figure \ref{fig_common_names_vs_cossim} and the threshold was calculated in the same way. It should be emphasised that both features produce an approximately equal similarity threshold.

The high similarity values of the contents of more than approx. 0.7 is clear to see. It seems that already one single common reference is sufficient to confirm the respective patent-publication pair. 
Zero common references cannot be considered as "no match" due to the poor data quality on the patent side.

\section*{Reduction of patent-publication pair ambiguity}
Following data preparation and feature creation, the last step in our workflow is to filter the identified raw publication-patent pairs in order to reduce the number of many-to-many relationships. This step involves, firstly, determining the patent classes (IPC) to be considered for filtering and, secondly, ranking the publication-patent connections using the features described above.

\subsection*{Filter by patent classes}
Allowing only certain patent classes for each scientific discipline is a common step in filtering matched publication-patent pairs. For example, \cite{Dornbusch2015Academic} manually match patent classes to Scopus science fields to ensure content-related correspondence. Since the approach presented here only covers items within the domain of "medicine", we introduce a statistical process to select permitted patent classes automatically. Submitted patents are classified by patent offices into different patent classes (IPC) based on their content. The system of IPC is based on the Strasbourg Agreement from 1971 (see e.g. \cite{wiki:strasbourg}) and allows classification into eight main sections and approx. 80k subdivisions. It has a tree structure consisting of Section, Class, Subclass and Group. A patent is usually assigned to several patent classes. Here we treat each class individually.
In this study, only the top three levels (Section, Class, Subclass) of an IPC are considered.

\begin{figure}[h!]\centering
\captionsetup{width=1\linewidth}
\includegraphics[width=1\linewidth]{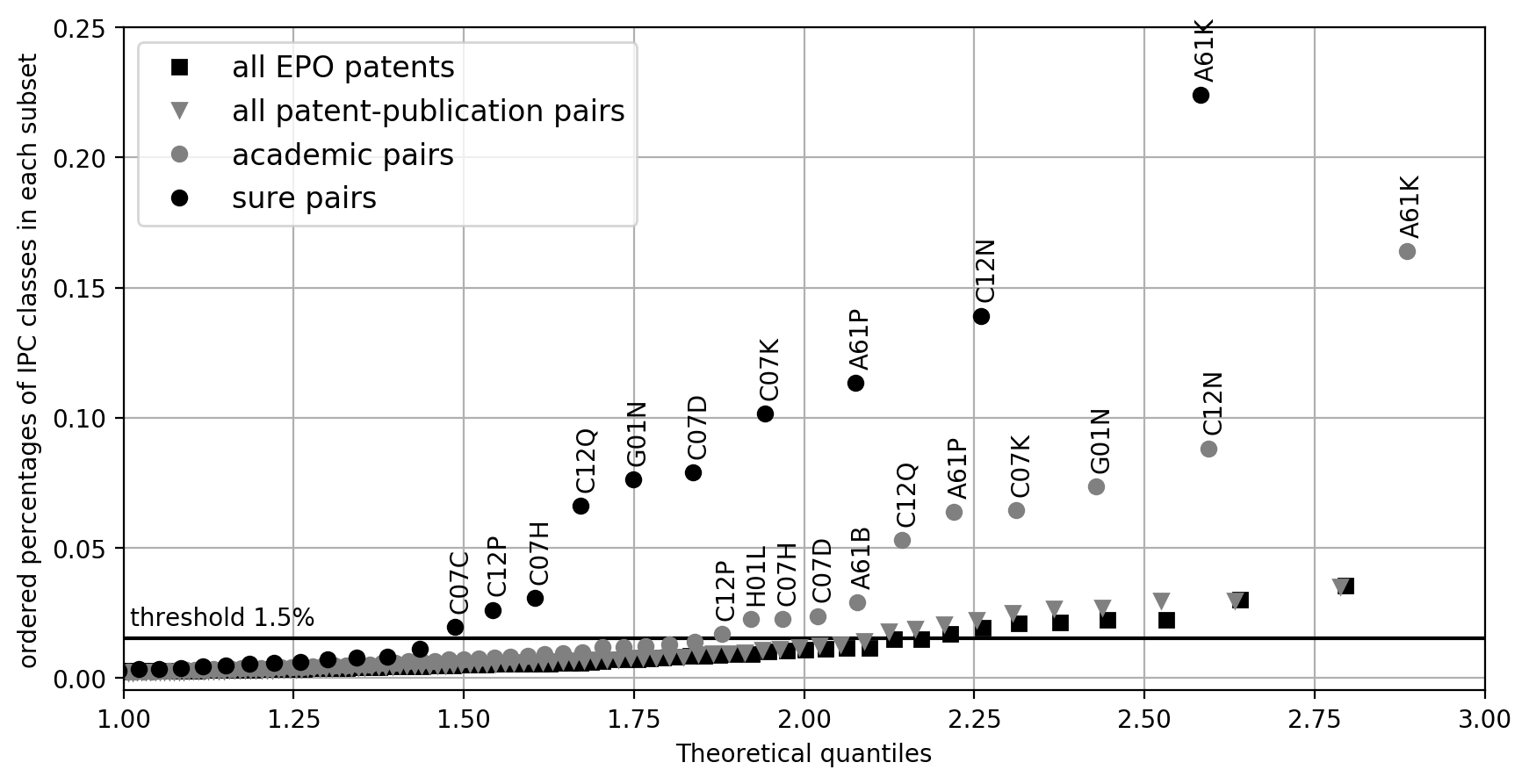}
\caption{Zoom into a Q-Q plot of the distribution of patent classes (IPC) for different subsets: (1) all EPO patent families with filing in the period of this study, (2) all identified patent-publication pairs, (3) academic patents indicated by 'prof' in inventor name or 'univ' in applicant name and (4) sure pairs with very strict criteria (at least 4 common names with same country or at least 4 common references or at least a cosine similarity of 0.95)
}
\label{fig_qq_diagram} 
\end{figure}

As a basic assumption, different scientific disciplines also differ in the composition of the patent classes of the resulting patents.
To find automatically allowed patent classes, the distributions of the patent classes in different subsets are examined. The distributions of the IPC are plotted in a common Q-Q plot (see Figure \ref{fig_qq_diagram}). The respective comparison of a subset with a normal distribution (which is presented in a Q-Q plot) also allows a comparison of subsets among each other.

The subsets used for comparison are:
\begin{itemize}
    \item all available patents from EPO in the period of this study (2000-2005). This subset is used as a baseline.
    \item all patent-publication pairs found by common names, no additional filtering.
    \item identifiable academic pairs as described above.
    \item for the subset "sure pairs" very strict restrictions were chosen. Much stricter than those used later in the ranking. Either 4 identical names and their countries must match OR 4 identical references must exist OR a cosine similarity of at least 0.95 must exist. The motto is to consider as few false patent-publication pairs as possible.
\end{itemize}

The comparison of the distributions (see Figure \ref{fig_qq_diagram}) shows that the distribution of the patent classes in identified patent-publication pairs without any further processing does not deviate significantly from the distribution of the baseline, i.e. all EPO patents in the period.

The "academic patents" and the "sure pairs", on the other hand, show a significant deviation from the baseline distribution. 

We set a percentage limit of 1.5$\%$ of the IPC classes in each subset as an arbitrary threshold. Above this level, the patent classes of the subsets "academic pairs" and "sure pairs" are qualitatively almost identical. In the following, the patent classes of the subset "sure pairs" that deviate from the baseline are used as filters.

\subsection*{Rank by cosine similarity and disambiguate features}
As a final step in the presented workflow, the identified patent-publication pairs are ranked.
For a patent-publication pair to be considered valid, one of the following conditions must be met:
\begin{itemize}
    \item the number of intersect names is at least 3
    \item the number of common references is at least 1
    \item cosine similarity is at least above the mean value of the two found thresholds by common names (see Figure \ref{fig_common_names_vs_cossim}) and by common references (see Figure \ref{fig_common_references_vs_cossim}). Only the best three that meet this condition are considered valid pairs. In addition, the similarity value is increased by 0.1 for an "academic patent".
\end{itemize}

\begin{figure}[h!]\centering
\captionsetup{width=1\linewidth}
\includegraphics[width=1\linewidth]{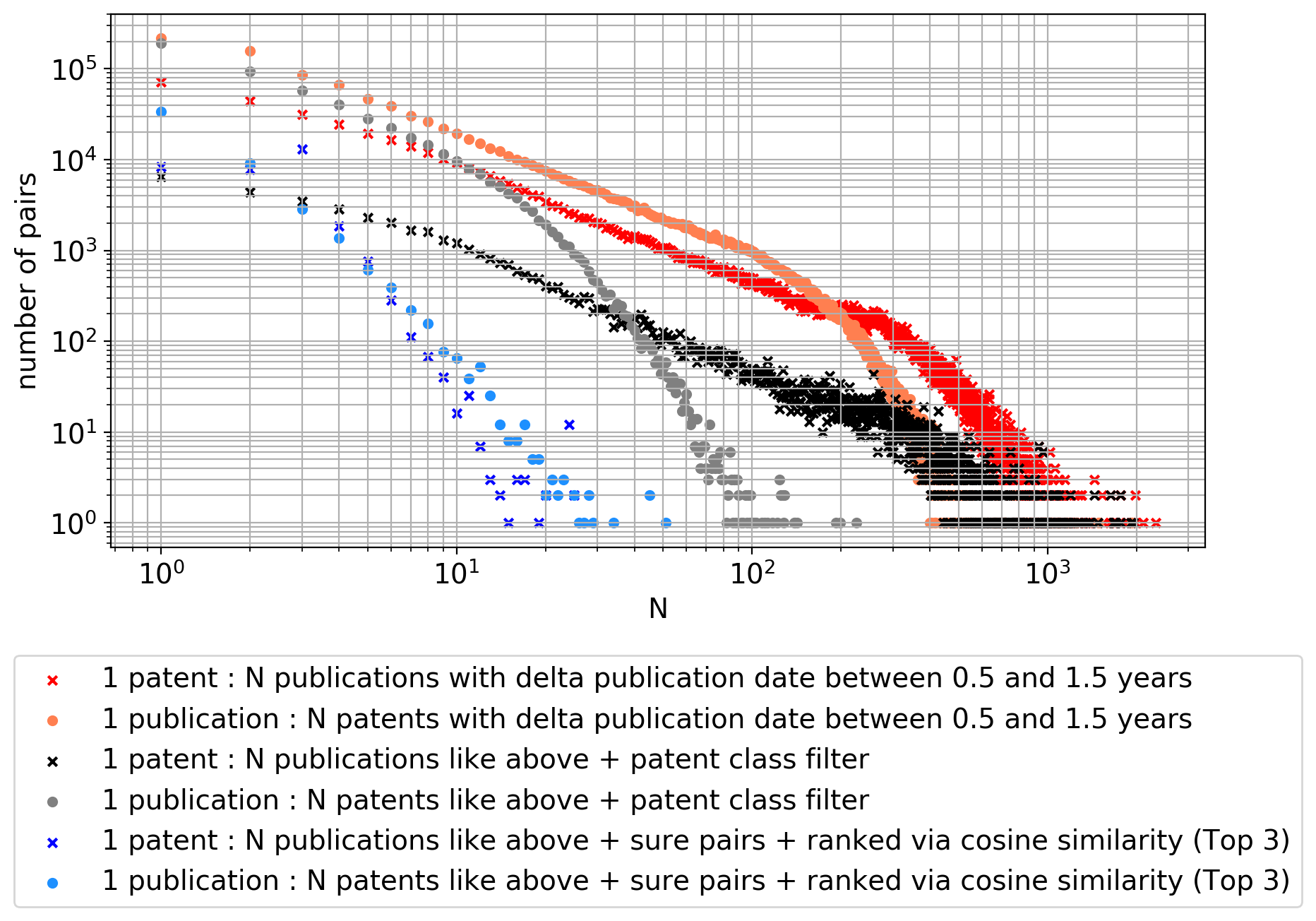}
\caption{Distribution of N:N relationships of patent-publication pairs for different filtering steps: (1) raw pairs in orange and red, (2) patent class filter in black and grey and (3) final results after ranking in blue and light blue.
}
\label{fig_NN_relationsships} 
\end{figure}

Figure \ref{fig_NN_relationsships} offers visual quality control of the reduction in many-to-many relationships from the raw pairs (orange and red), through filtering using patent classes (black and grey), to the final result after ranking (blue and light blue). 

Since many-to-many relationships are very difficult to present, 1:N relationships are shown in each case: (1) one patent links to many publications (crosses in Figure \ref{fig_NN_relationsships} ) and (2) one publication links to many patents (circles in Figure \ref{fig_NN_relationsships}).
The completely unrealistic 1 to $>$ 1000 relationships from the raw pairs are efficiently reduced by the IPC filter; by applying the similarity ranking this selection was limited to the most similar documents. Nevertheless, patent-publication pairs characterised by a high number of overlapping names or references were not discarded, but included.

\section*{Evaluation}
\begin{figure}[h!]\centering
\captionsetup{width=1\linewidth}
\includegraphics[width=1\linewidth]{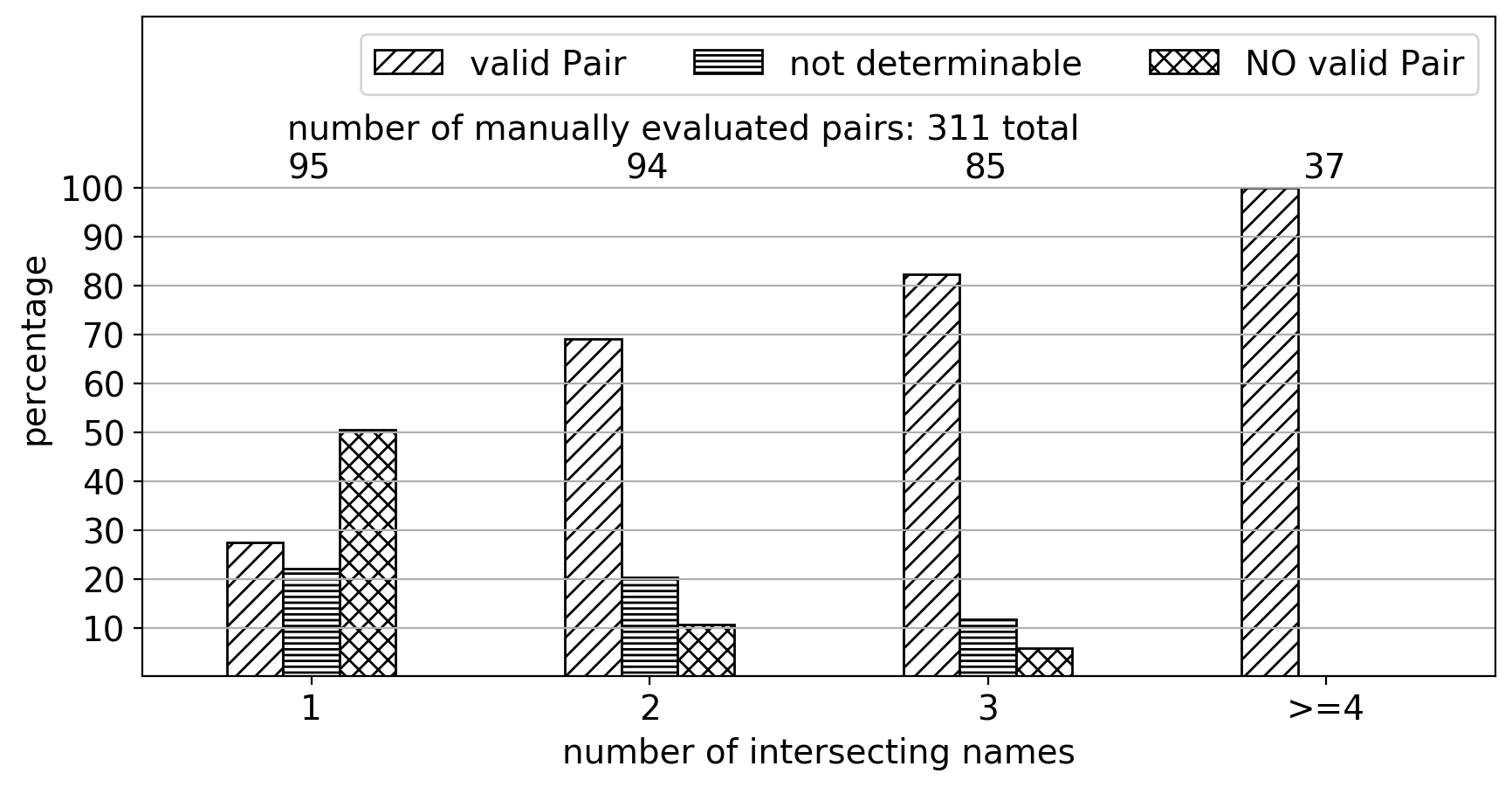}
\caption{Result of a manual review of the validity of a random selection of the found publication-patent pairs. }
\label{fig_evaluation}
\end{figure}

As mentioned in the introduction, an evaluation using the KEINS database is not possible in our case, and we are not aware of any other freely available data on (European) publication-patent pairs. To enable some form of evaluation to be conducted, we therefore developed a small web service. This tool allows a manual review to be performed by presenting the required information to a human reviewer. The results can be seen in Figure \ref{fig_evaluation}.
Starting with the found publication-patent pairs, a defined quantity of pairs per number of common names was randomly selected and evaluated.  By referring to the publication metadata and abstracts and the complete patent documents, we checked the robustness of these pairings individually and thus classified them into three categories: "valid pair", "no valid pair" and "not determinable". We are aware that it would take at least one specialist in each subject area to be sure of performing this classification correctly in every instance; hence the use of the "not determinable" category in case of doubt. Even so, this evaluation still succeeds in providing a certain degree of quality control for the new filter features presented in this work.
\\
With only one author/inventor name in common, about half of the found pairs are clearly 
identifiable as false despite the applied filters. Nonetheless, this percentage is even higher without the applied filters, as illustrated by the high number of ambiguities for raw publication-patent pairs shown in Figure \ref{fig_NN_relationsships}. As soon as the number of common names rises to two, the percentage of pairs clearly identifiable as false drops to about 10$\%$. And with four or more identical names, the number of clearly invalid pairs in the results falls to zero.

\section*{Discussion and outlook}
In this study, we reproduced existing workflows to match patents and publications in the life sciences domain over an example five-year period using only publicly available data sets. In a major refinement to this approach, we then applied and examined two new filter features designed to tackle the problem of ambiguity: document cosine similarity using word embeddings, and the identification of common references in the two document types. In addition, we presented an automated method of filtering permitted patent classes. \
If we had one wish for the data records, it would be the unique identification of authors/inventors via a system such as ORCID. Unfortunately, this is not mandatory for either patents or publications. Thus, the ambiguity of identical names remains, and the only solution is to filter the results using additional features.

All the SQL and Python codes of the workflow and the resulting patent-publication pairs are available on GitHub under a permissive open source licence. As part of a joint research project (QuaMedFo), this workflow was adapted to the publication data provided by the pilot faculties of the joint research project, and thus also the timeframe is different to what is presented here. Individual pairs were aggregated to medical specialties, providing comparability among the three pilot faculties. Detailed results are presented in the anthology accompanying the QuaMedFo project (\cite{quamedfo_sammelband}).

As the next step and logical consequence of the present study, steps are currently being taken to implement the patent-publication matching workflow into LIVIVO (\cite{livivo}), ZB MED's Search Portal for Life Sciences. By extending the time range to all available patent data sets from the European Patent Office and connecting this data to the more than 70$\cdot 10^6$ publications available in LIVIVO, we will be able to offer our users additional valuable information on the links between the life sciences and the economy; for example, \cite{seeberfr} mentions that additional data are often exclusively published in patents.

\section*{Acknowledgements}
The authors would like to express their gratitude for the input and constructive feedback provided by colleagues from the other institutions involved in the QuaMedFo Project: the German Centre for Higher Education Research and Science Studies (DZHW), the Leibniz Information Centre for Economics (ZBW) and the University Medical Faculty in Göttingen (UMG). 

We also wish to extend our thanks to the proofreaders at ZB MED, our colleagues A. Halder and  E. Seidlmayer.

\section*{Declarations}
\subsubsection*{Funding}
The present work is part of the research project "QuaMedFo" ("Qualitätsmaße zur Evaluierung medizinischer Forschung" -- "Quality measures for the evaluation of medical research") in the funding line "Quantitative Science Research" of the German BMBF (German Federal Ministry of Education and Research). Its funding reference number is 16PU17011C.

\subsubsection*{Conflict of interest}
None of the authors have conflicts of
interest in the subject matter or materials discussed in this manuscript.

\bibliographystyle{plainnat}
\bibliography{references}

\begin{thebibliography}{32}
\providecommand{\natexlab}[1]{#1}
\providecommand{\url}[1]{\texttt{#1}}
\expandafter\ifx\csname urlstyle\endcsname\relax
  \providecommand{\doi}[1]{doi: #1}\else
  \providecommand{\doi}{doi: \begingroup \urlstyle{rm}\Url}\fi

\bibitem[Agrawal and Henderson(2002)]{mit1}
Ajay Agrawal and Rebecca Henderson.
\newblock Putting patents in context: Exploring knowledge transfer from {MIT}.
\newblock \emph{Management Science}, 48(1):\penalty0 44--60, 2002.
\newblock \doi{10.1287/mnsc.48.1.44.14279}.

\bibitem[Baroni et~al.(2014)Baroni, Dinu, and Kruszewski]{dontcountpredict}
Marco Baroni, Georgiana Dinu, and Germán Kruszewski.
\newblock Don't count, predict! a systematic comparison of context-counting vs.
  context-predicting semantic vectors.
\newblock volume~1, pages 238--247, 06 2014.
\newblock \doi{10.3115/v1/P14-1023}.

\bibitem[Brockmann(2017)]{Brockmann2017Global}
Dirk Brockmann.
\newblock Global connectivity and the spread of infectious diseases.
\newblock In \emph{Nova Acta Leopoldina}, number 419. Robert Koch-Institut,
  2017.
\newblock \doi{http://dx.doi.org/10.25646/2797}.

\bibitem[Chen et~al.(2021)Chen, Hu, Jimenez-Ruiz, Holter, Antonyrajah, and
  Horrocks]{owl2vec}
Jiaoyan Chen, Pan Hu, Ernesto Jimenez-Ruiz, Ole~Magnus Holter, Denvar
  Antonyrajah, and Ian Horrocks.
\newblock Owl2vec*: embedding of owl ontologies.
\newblock \emph{Machine Learning}, 2021.
\newblock \doi{10.1007/s10994-021-05997-6}.

\bibitem[Dahlborg et~al.(2013)Dahlborg, Lewensohn, and Sundberg]{sweden1}
Charlotta Dahlborg, Danielle Lewensohn, and Carl~Johan Sundberg.
\newblock Investigating inventive productivity at sweden’s largest medical
  university.
\newblock \emph{International Journal of Technology Transfer and
  Commercialisation}, 12:\penalty0 102--120, 01 2013.
\newblock \doi{10.1504/IJTTC.2013.064154}.

\bibitem[{Deutsches Institut für Medizinische Dokumentation und Information
  (DIMDI)}(2019)]{mesh_de}
{Deutsches Institut für Medizinische Dokumentation und Information (DIMDI)}.
\newblock German mesh, 2019.
\newblock URL
  \url{https://www.dimdi.de/dynamic/de/klassifikationen/weitere-klassifikationen-und-standards/mesh/}.
\newblock [Online; accessed 1-September-2020].

\bibitem[Devlin et~al.(2018)Devlin, Chang, Lee, and Toutanova]{bert}
Jacob Devlin, Ming-Wei Chang, Kenton Lee, and Kristina Toutanova.
\newblock Bert: Pre-training of deep bidirectional transformers for language
  understanding.
\newblock 2018.
\newblock \doi{10.48550/ARXIV.1810.04805}.
\newblock URL \url{https://arxiv.org/abs/1810.04805}.

\bibitem[Dornbusch and Neuh\"{a}usler(2015)]{Dornbusch2015Academic}
Friedrich Dornbusch and Peter Neuh\"{a}usler.
\newblock Academic patents in germany.
\newblock Studien zum deutschen Innovationssystem 6-2015, Berlin, 2015.
\newblock URL \url{http://hdl.handle.net/10419/156616}.

\bibitem[Ferreira et~al.(2012)Ferreira, Gonçalves, and Laender]{ferreira1}
Anderson Ferreira, Marcos Gonçalves, and Alberto Laender.
\newblock A brief survey of automatic methods for author name disambiguation.
\newblock \emph{ACM SIGMOD Record}, 41:\penalty0 15--26, 08 2012.
\newblock \doi{10.1145/2350036.2350040}.

\bibitem[Grover and Leskovec(2016)]{node2vec}
A.~Grover and J.~Leskovec.
\newblock node2vec: Scalable feature learning for networks.
\newblock In \emph{International Conference on Knowledge Discovery and Data
  Mining (KDD)}, 2016.

\bibitem[Gurulingappa et~al.(2010)Gurulingappa, M{\"{u}}ller, Klinger,
  Mevissen, Hofmann{-}Apitius, Friedrich, and Fluck]{berndpaper}
Harsha Gurulingappa, Bernd M{\"{u}}ller, Roman Klinger, Heinz{-}Theodor
  Mevissen, Martin Hofmann{-}Apitius, Christoph~M. Friedrich, and Juliane
  Fluck.
\newblock Prior art search in chemistry patents based on semantic concepts and
  co-citation analysis.
\newblock In Ellen~M. Voorhees and Lori~P. Buckland, editors, \emph{Proceedings
  of The Nineteenth Text REtrieval Conference, {TREC} 2010, Gaithersburg,
  Maryland, USA, November 16-19, 2010}, volume 500-294 of \emph{{NIST} Special
  Publication}. National Institute of Standards and Technology {(NIST)}, 2010.
\newblock URL
  \url{http://trec.nist.gov/pubs/trec19/papers/fraunhofer-scai.chem.rev.pdf}.

\bibitem[{Institut national de la santé et de la recherche médicale
  (INSERM)}(2019)]{mesh_fr}
{Institut national de la santé et de la recherche médicale (INSERM)}.
\newblock Le mesh bilingue anglais - français, 2019.
\newblock URL \url{http://mesh.inserm.fr/FrenchMesh/}.
\newblock [Online; accessed 20-April-2022].

\bibitem[Kim et~al.(2021)Kim, Kim, and Owen‐Smith]{jinseok1}
Jinseok Kim, Jenna Kim, and Jason Owen‐Smith.
\newblock Ethnicity‐based name partitioning for author name disambiguation
  using supervised machine learning.
\newblock \emph{Journal of the Association for Information Science and
  Technology}, 72, 02 2021.
\newblock \doi{10.1002/asi.24459}.

\bibitem[Lee et~al.(2019)Lee, Yoon, Kim, Kim, Kim, So, and Kang]{Lee_2019}
Jinhyuk Lee, Wonjin Yoon, Sungdong Kim, Donghyeon Kim, Sunkyu Kim, Chan~Ho So,
  and Jaewoo Kang.
\newblock {BioBERT}: a pre-trained biomedical language representation model for
  biomedical text mining.
\newblock \emph{Bioinformatics}, 36\penalty0 (4):\penalty0 1234--1240, sep
  2019.
\newblock \doi{10.1093/bioinformatics/btz682}.

\bibitem[Lemke et~al.(2022)Lemke, Witthake, and Peters]{zbw_altmetrics}
Steffen Lemke, Anne Witthake, and Isabella Peters.
\newblock Altmetrics for german medical research: what leads to research
  articles achieving policy impact?
\newblock In \emph{26th International Conference on Science, Technology and
  Innovation Indicators (STI 2022)}, 2022.
\newblock \doi{10.5281/zenodo.6645109}.

\bibitem[Lippert and F{\"o}rstner(2024)]{quamedfo_sammelband}
Klaus Lippert and Konrad~U. F{\"o}rstner.
\newblock \emph{Nutzung von Patentdaten zur Erfassung der wirtschaftlichen
  Verwertung von Forschung}, pages 97--107.
\newblock Springer Fachmedien Wiesbaden, Wiesbaden, 2024.
\newblock ISBN 978-3-658-43683-4.
\newblock \doi{10.1007/978-3-658-43683-4_6}.
\newblock URL \url{https://doi.org/10.1007/978-3-658-43683-4_6}.

\bibitem[Lissoni et~al.(2008)Lissoni, Llerena, Mckelvey, and
  Sanditov]{lissoni1}
Francesco Lissoni, Patrick Llerena, Maureen Mckelvey, and Bulat Sanditov.
\newblock Academic patenting in europe: New evidence from the keins database.
\newblock \emph{Research Evaluation}, 17, 02 2008.
\newblock \doi{10.3152/095820208X287171}.

\bibitem[Lissoni et~al.(2009)Lissoni, Lotz, Schovsbo, and Treccani]{provprivi2}
Francesco Lissoni, Peter Lotz, Jens Schovsbo, and Adele Treccani.
\newblock Academic patenting and the professor's privilege: Evidence on denmark
  from the keins database.
\newblock \emph{Science and Public Policy - SCI PUBLIC POLICY}, 36:\penalty0
  595--607, 10 2009.
\newblock \doi{10.3152/030234209X475443}.

\bibitem[Milojevic(2013)]{Milojevic2013AccuracyOS}
Stasa Milojevic.
\newblock Accuracy of simple, initials-based methods for author name
  disambiguation.
\newblock \emph{J. Informetrics}, 7:\penalty0 767--773, 2013.
\newblock \doi{10.1016/j.joi.2013.06.006}.

\bibitem[Müller et~al.(2017)Müller, Poley, Pössel, Hagelstein, and
  Gübitz]{livivo}
Bernd Müller, Christoph Poley, Jana Pössel, Alexandra Hagelstein, and Thomas
  Gübitz.
\newblock Livivo – the vertical search engine for life sciences.
\newblock \emph{Datenbank-Spektrum}, 17:\penalty0 29--34, 2017.
\newblock \doi{10.1007/s13222-016-0245-2}.
\newblock URL \url{https://www.livivo.de}.

\bibitem[{National Institute of Standards and Technology
  (NIST)}(2022)]{cosdist}
{National Institute of Standards and Technology (NIST)}.
\newblock Cosine distance, cosine similarity, angular cosine distance, angular
  cosine similarity, 2022.
\newblock URL
  \url{https://www.itl.nist.gov/div898/software/dataplot/refman2/auxillar/cosdist.htm}.
\newblock [Online; accessed 20-April-2022].

\bibitem[{National Library of Medicine (US) (NLM)}(2000)]{pmc}
{National Library of Medicine (US) (NLM)}.
\newblock Pubmed central, 2000.
\newblock URL \url{https://www.ncbi.nlm.nih.gov/pmc/}.
\newblock [Online; accessed 20-April-2022].

\bibitem[{National Library of Medicine (US) (NLM)}(2022)]{mesh_en}
{National Library of Medicine (US) (NLM)}.
\newblock {Medical Subject Headings MeSH}, 2022.
\newblock URL \url{https://www.nlm.nih.gov/mesh/meshhome.html}.
\newblock [Online; accessed 20-April-2022].

\bibitem[Ristoski and Paulheim(2016)]{rdf2vec}
Petar Ristoski and Heiko Paulheim.
\newblock {RDF2Vec: RDF Graph Embeddings for Data Mining}.
\newblock In \emph{The Semantic Web – ISWC 2016}. Springer International
  Publishing, 2016.

\bibitem[Seeber(2007)]{seeberfr}
Frank Seeber.
\newblock Patent searches as a complement to literature searches in the life
  sciences—a ‘how-to’ tutorial.
\newblock \emph{Nature protocols}, 2:\penalty0 2418--28, 02 2007.
\newblock \doi{10.1038/nprot.2007.355}.

\bibitem[{van Dongen} et~al.(2014){van Dongen}, Winnink, and
  Tijssen]{VANDONGEN201427}
Peter {van Dongen}, Jos Winnink, and Robert Tijssen.
\newblock Academic inventions and patents in the netherlands: A case study on
  business sector exploitation.
\newblock \emph{World Patent Information}, 38:\penalty0 27--32, 2014.
\newblock ISSN 0172-2190.
\newblock \doi{https://doi.org/10.1016/j.wpi.2014.03.002}.
\newblock URL
  \url{https://www.sciencedirect.com/science/article/pii/S0172219014000404}.

\bibitem[Verberne et~al.(2019)Verberne, Chios, and
  Wang]{Verberne2019ExtractingAM}
Suzan Verberne, Ioannis Chios, and Jian Wang.
\newblock Extracting and matching patent in-text references to scientific
  publications.
\newblock In \emph{BIRNDL@SIGIR}, 2019.

\bibitem[Von~Proff et~al.(2012)Von~Proff, Buenstorf, and Hummel]{profprivi}
Sidonia Von~Proff, Guido Buenstorf, and Martin Hummel.
\newblock University patenting in germany before and after 2002: what role did
  the professors' privilege play?
\newblock \emph{Industry and Innovation}, 19\penalty0 (1):\penalty0 23--44,
  2012.
\newblock \doi{10.1080/13662716.2012.649060}.

\bibitem[Voskuil and Verberne(2021)]{DBLP_journals/corr/abs-2101-01039}
Ken~S. Voskuil and Suzan Verberne.
\newblock Improving reference mining in patents with {BERT}.
\newblock \emph{CoRR}, abs/2101.01039, 2021.
\newblock URL \url{https://arxiv.org/abs/2101.01039}.

\bibitem[{Wikipedia}(2022{\natexlab{a}})]{wiki:hochschullehrerprivileg}
{Wikipedia}.
\newblock Hochschullehrerprivileg --- {W}ikipedia{,} the free encyclopedia,
  2022{\natexlab{a}}.
\newblock URL \url{https://de.wikipedia.org/wiki/Hochschullehrerprivileg}.
\newblock [Online; accessed 20-April-2022].

\bibitem[{Wikipedia}(2022{\natexlab{b}})]{wiki:strasbourg}
{Wikipedia}.
\newblock Strasbourg agreement --- {W}ikipedia{,} the free encyclopedia,
  2022{\natexlab{b}}.
\newblock URL \url{https://en.wikipedia.org/wiki/
  Strasbourg_Agreement_Concerning_the_International_Patent_Classification}.
\newblock [Online; accessed 20-April-2022].

\bibitem[Wilkinson et~al.(2016)Wilkinson, Dumontier, and Aalbersberg]{fair1}
M.~Wilkinson, M.~Dumontier, and I.~Aalbersberg.
\newblock The fair guiding principles for scientific data management and
  stewardship.
\newblock \emph{Sci Data}, 2016.
\newblock \doi{10.1038/sdata.2016.18}.

\end{thebibliography}



\end{document}